\documentclass[english,aps,prb,superscriptaddress,longbibliography]{revtex4-2}
\usepackage{babel}
\usepackage{amsmath,mathtools}
\usepackage{amssymb}
\usepackage{graphicx}
\usepackage{tabularx}
\usepackage{lipsum}
\usepackage{xcolor}
\usepackage{soul}

\begin{document}
\title{Topological Zero Modes in Non-Hermitian Topolectrical Systems: Size and Impedance Control }
\author{S M Rafi-Ul-Islam}
\email{rafiul.islam@u.nus.edu}
\selectlanguage{english}
\affiliation{Department of Electrical and Computer Engineering, National University of Singapore, Singapore}
\author{Zhuo Bin Siu}
\email{elesiuz@nus.edu.sg}
\selectlanguage{english}
\affiliation{Department of Electrical and Computer Engineering, National University of Singapore, Singapore}
\author{Md. Saddam Hossain Razo}
\email{shrazo@u.nus.edu}
\affiliation{Department of Electrical and Computer Engineering, National University of Singapore, Singapore 117583, Republic of Singapore}
\author{Mansoor B.A. Jalil}
\email{elembaj@nus.edu.sg}
\selectlanguage{english}
\affiliation{Department of Electrical and Computer Engineering, National University of Singapore, Singapore}

\begin{abstract}
Lattice models that exhibit topological zero-energy boundary modes show a remarkable sensitivity to the systems length and the applied boundary conditions. We present the exact solutions for the eigenenergies and band gap of the topological modes in a generalized two-band non-Hermitian model on topolectrical circuit platform, which show an extra-ordinary dependence on the system size. Additionally, we show that the introduction of non-Hermiticity results in the recovery of topological modes with the exact eigenenergy of zero at a certain critical system size. Such (nearly) zero-admittance edge states are reflected in the large impedance peaks in the impedance spectrum. Our results reveal that the boundary modes of a non-Hermitian lattice can be tuned via system size.
\end{abstract}
\maketitle

\section{Introduction}

The exploration of topological states of matter \cite{wang2017topological,zhang2008topological,sahin2024protected,rafi2020strain,fujita2011gauge,rafi2020topoelectrical,haldane2017nobel,rafi2023valley,rafi2020realization} has become an intriguing and promising area in condensed matter physics \cite{trebin1982topology,sun2020spin,rafi2023conductance,laflorencie2016quantum,sun2019field,irkhin2019modern} that provides the promise of electronic properties that are resilient against local perturbations \cite{groning2018engineering,rafi2020anti,chen2012robustness}. A pivotal aspect of this exploration lies in the investigation of topological boundary modes \cite{kane2014topological,nayak2021evidence,borgnia2020non}, which play a crucial role not only in Hermitian systems \cite{obana2019topological,rafi2022valley,nakada1996edge,hafezi2013imaging} but also present both challenges and opportunities \cite{butler2013progress,kou2017two,xie2022progress,yang2023topological} in the nuanced realm of non-Hermitian systems \cite{ashida2020non,bergholtz2021exceptional,rafi2024knots,kawabata2019symmetry,rafi2024twisted,siu2023terminal,gong2018topological}. Various topological boundary states, such as topological zero-modes \cite{koshino2014topological,rafi2022system,song2019breakup,noh2020braiding}, corner modes \cite{imhof2018topolectrical,zhu2020distinguishing,proctor2020robustness,rafi2022type}, and hinge states \cite{ghorashi2019second,fu2021chiral,rafi2024chiral}, have been manifested in diverse systems spanning topolectrical (TE) circuits \cite{helbig2020generalized,rafi2022system,zhang2023anomalous,zou2021observation,rafi2021non,sahin2023impedance,hofmann2020reciprocal,rafi2021topological}, photonics \cite{lu2014topological,hafezi2013imaging}, optics \cite{barik2018topological,soskin2016singular}, superconducting setups \cite{qi2011topological,zhang2019multiple}, and metamaterials \cite{krishnamoorthy2012topological,liu2017disorder}. TE circuits, in particular, stand out owing to their accessibility and ease of implementation \cite{imhof2018topolectrical}. Our work builds on prior studies of non-Hermitian systems, such as Ref. \cite{song2019breakup}, by focusing on the size-dependent recovery of topological zero modes (TZMs) in finite non-Hermitian SSH chains with asymmetric coupling. Unlike previous works that explore bulk properties, we derive exact solutions for eigenenergies and band gaps (Eqs. \ref{req4}, \ref{req5}), highlighting the role of asymmetric coupling in tuning TZMs at a critical system size \(M_c\) (Eq. \ref{req6}), offering new insights for circuit-based topological phenomena.

Topological zero modes (TZMs), which are localized at the edges or corners of a lattice in a topological system, possess a unique resilience due to the topological properties of the system. Despite the name ``topological zero modes," such modes do not occur exactly at zero eigenenergy in finite-sized Hermitian systems with boundaries but instead have eigenenergies that are sensitive to the system size. Studying the dislocation of TZMs from zero eigenenergy in finite lattices is important for understanding the behavior of these modes in real-world systems, in which the lattice size is finite, or in non-Hermitian systems. In non-Hermitian SSH models with asymmetric coupling, the non-Hermitian skin effect (NHSE) can lead to the localization of eigenstates at one boundary \cite{PhysRevResearch.4.013243,gggh-jy6j,rafi2025critical,rafi2022interfacial}, potentially coexisting with TZMs. While our study focuses on TZM recovery, the NHSE may influence mode localization, enhancing their robustness for practical applications. For example, in TE realizations of such systems, the inevitable presence of series resistances in real-world components introduces some degree of non-Hermiticity into such systems. Such imperfections in real-world topological systems frequently cause additional deviations of the topological boundary modes from the precisely predicted zero-energy states. In particular, the dislocation of TZMs in finite lattice sizes can have important consequences in the implementation of quantum algorithms, in which the stability and robustness of the TZM modes is essential for the proper functioning of the algorithm \cite{sarma2015majorana,bomantara2018quantum}. Moreover, 1D topological structures, such as those studied here, show promise for applications like wireless power transfer (WPT) and polarization conversion. For instance, 1D photonic topological insulators using split ring resonators enable efficient, disorder-immune WPT for medical implants and electronics \cite{Wang2024}, while topological edge states in dimer chains facilitate robust polarization conversion for optical communication \cite{Yang2025}, and impedance-based switching, and frequency-selective filtering. Our non-Hermitian TE circuits, with tunable TZMs and large impedance peaks (Fig. \ref{fig3}d–f), could enable compact, reconfigurable designs for such applications.

In this letter, we provide a clear understanding of the size-dependent characteristics of the boundary states in a non-Hermitian system implemented in a topolectrical circuit under open boundary conditions. We derive the exact solutions for the edge states of the non-Hermitian model with non-reciprocal coupling and onsite loss terms to explore the effect of a finite system size. Our analytical results reveal how the system size affects the boundary states localization. In short chains, the interaction of the edge modes with each other via near-field coupling effect causes the boundary states to deviate significantly from zero energy \cite{guo2021sensitivity,song2019breakup}. Furthermore, a tunable resistor-induced staggered gain and loss term on the sublattice sites shifts the energies of the topological modes towards zero at parameter regions closer to the ideal infinite-size value in short chains compared to the corresponding Hermitian limit without the loss and gain terms. Moreover, we find that the impedance spectra shows a large peak only at a critical size where the band-gap between boundary modes is close to zero in Hermitian systems and exactly zero in non-Hermitian ones. By incorporating an additional common grounding negative intercell capacitor via a negative impedance converter (NIC) at each node, the resonant frequency becomes independent of intercell capacitor, allowing tunable edge states via intercell capacitor variations at a fixed frequency, enhancing practical control over topological properties. Our results demonstrate the novel phenomena associated with the boundary modes can be characterized exactly in a finite non-Hermitian system.

\section{Results}
\begin{figure*}[htp!]
  \centering
    \includegraphics[width=0.8\textwidth]{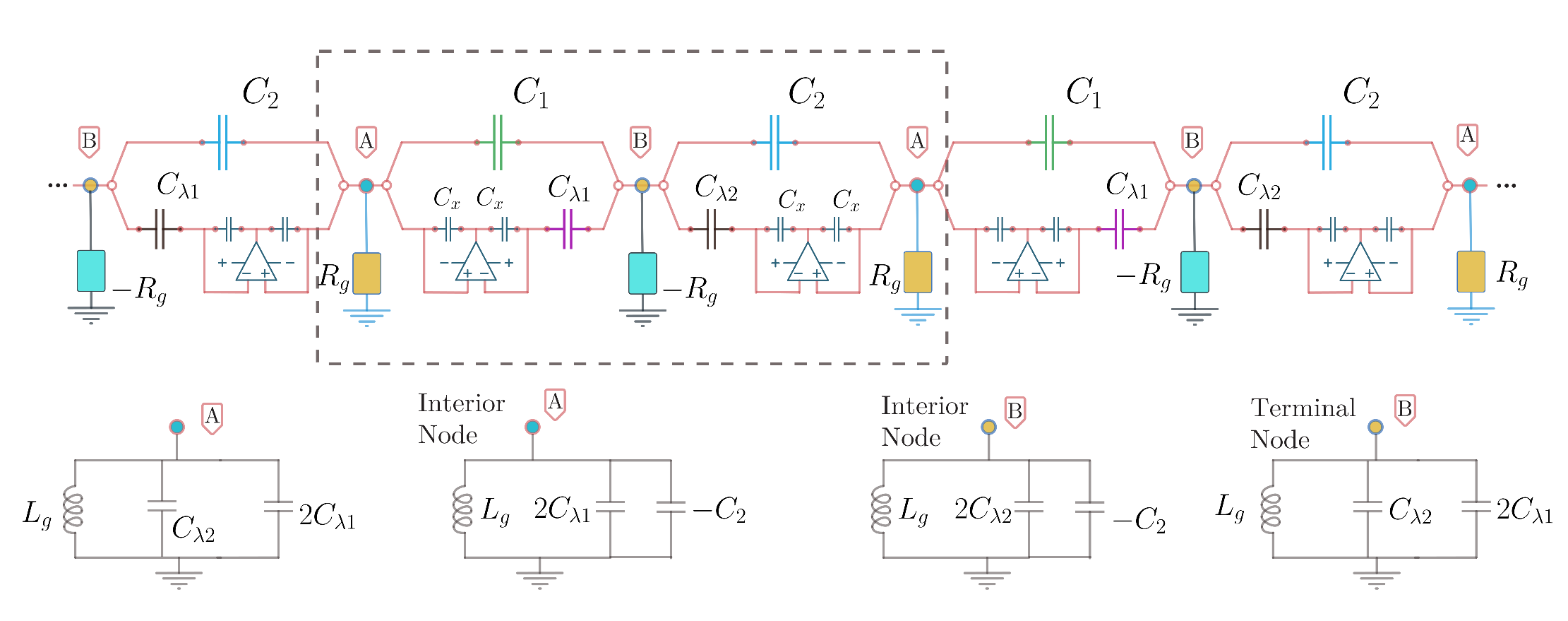}
  \caption{Circuit model of a generalized non-Hermitian SSH chain comprising A and B sublattice nodes. Non-reciprocal intra- and inter-cell coupling is represented by capacitors \( C_1 \pm C_{\lambda 1} \) and \( C_2 \pm C_{\lambda 2} \), where the asymmetry \( \pm C_{\lambda i} \) (with \( i=1,2 \)) is induced through a negative impedance converter (NIC) for current inversion. A common inductor \( L_g \) adjusts the onsite potential, setting the driving frequency at resonance to enable an equivalent tight-binding model representation. Gain and loss terms applied to A and B nodes, respectively, are controlled via a tunable resistor (\( R_g \)), with or without NIC elements to preserve resistor sign. An additional grounding capacitor \(-C_2\) via NIC is connected to each node to modify the onsite potential, making the resonant frequency \( f_r = \frac{1}{2 \pi \sqrt{L_g (C_1 + C_{\lambda 1} + C_{\lambda 2})}} \) independent of \( C_2 \). Under open boundary conditions, additional grounding capacitances (\( C_2 \) and \( C_{\lambda 2} \)) maintain a consistent diagonal term in the circuit Laplacian matrix, facilitating a uniform shift in the admittance energy dispersion. The resonance frequency is applied to set the uniform shift to zero. Here, \(\omega\) denotes the angular frequency (\(\mathrm{rad/s}\)), and \(f_r\) is the frequency (\(\mathrm{Hz}\)), related by \(\omega = 2\pi f_r\).}
  \label{fig1}
\end{figure*}
We first consider a generalized non-Hermitian SSH chain comprising A and B sublattice nodes. The topolectrical circuit model and its connections are shown in Fig. \ref{fig1}. The capacitor values (\(C_1 \pm C_{\lambda 1}\)) and (\(C_2 \pm C_{\lambda 2}\)) denote the non-reciprocal intra and inter-cell coupling, respectively. The coupling asymmetry (\(\pm C_{\lambda i} \) with \(i=1,2\)) in the forward and backward direction is realized via a negative impedance converter (NIC) at current inversion \cite{imhof2018topolectrical,rafi2021topological}. An additional grounding capacitor \(-C_2\) via NIC is connected to each node, modifying the onsite potential. The common onsite potential is adjusted via a common inductor \(L_g\), which sets the resonance frequency of the driving alternating current at resonance so that the circuit can be treated as an equivalent tight-binding model (see Methods section for details). Furthermore, all the A and B type nodes are connected to onsite gain and loss terms, respectively. The modulation of the gain and loss terms in our TE circuit chain is realized by means of a tunable resistor without and with NIC component that preserves the sign of the resistor (\(R_g\)). In all equations, \(\omega\) represents the angular frequency (\(\mathrm{rad/s}\)), and \(f\) denotes the frequency (\(\mathrm{Hz}\)), with \(\omega = 2\pi f\). The circuit Laplacian, or admittance matrix, of an electrical network is defined by the relation \( \mathcal{J}(\omega) = I(\omega) V(\omega)^{-1} \), where \( I(\omega) \) and \( V(\omega) \) are the column current and voltage matrices, respectively. For the circuit array shown in Fig. \ref{fig1}, the admittance matrix under periodic boundary conditions (PBC) is given by

\begin{equation}
\mathcal{J}(\omega) = -i \omega \left( \mu \mathcal{I}_2 - \frac{1}{i \omega R_g} \sigma_z + \bar{\mathcal{J}}(\omega) \right).
\label{eqLap}
\end{equation}

In the above, \( \mu = \frac{1}{\omega^2 L_g} - (C_1 + C_{\lambda 1} + C_{\lambda 2}) \), and

\[
\bar{\mathcal{J}}(\omega) = (C_1 + C_2 \cos k_x + C_{\lambda 2} \cos k_x ) \sigma_x - (i C_{\lambda 1} - C_2 \sin k_x + C_{\lambda 2} \sin k_x) \sigma_y + i \frac{1}{\omega R_g} \sigma_z,
\]

where \( \sigma_i \) denotes the \(i\)-th Pauli matrix. The term \( \mu \) in Eq. \ref{eqLap} represents the global shift in the admittance dispersion. The common grounding inductor \( L_g \) is chosen to satisfy

\[
L_g = \left( \omega^2 (C_1 + C_{\lambda 1} + C_{\lambda 2}) \right)^{-1}
\]

in order to cancel the global shift in the admittance eigenspectrum. Consequently, the resonant frequency of the alternating signal is set at

\[
f_r = \frac{1}{2 \pi \sqrt{L_g (C_1 + C_{\lambda 1} + C_{\lambda 2})}}.
\]

\begin{figure*}[htp!]
  \centering
    \includegraphics[width=0.8\textwidth]{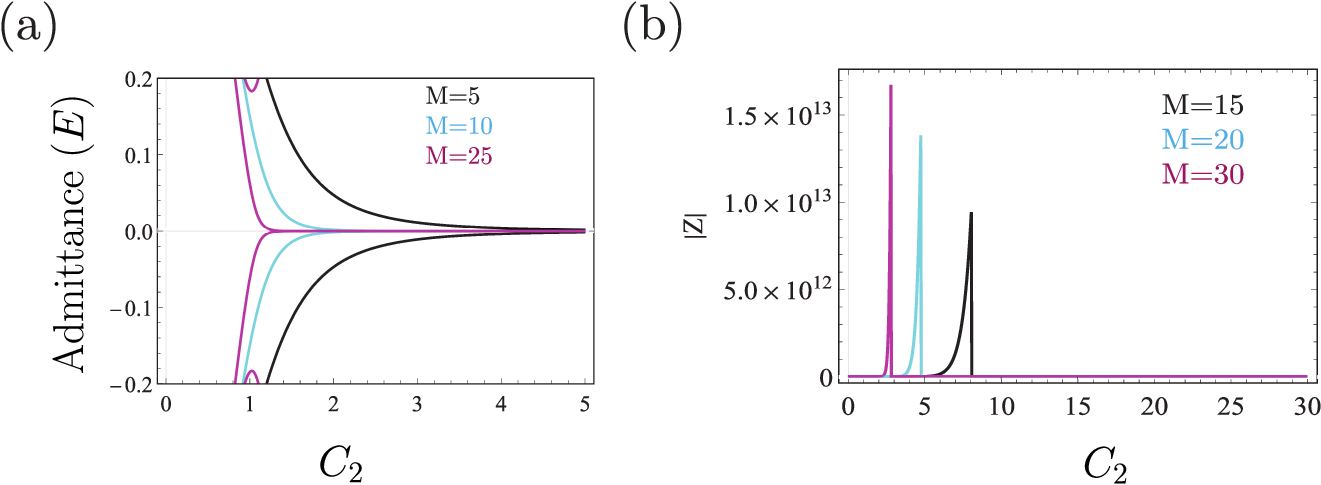}
  \caption{Size-dependent topological edge states in a Hermitian finite chain. (a) Admittance splitting of the two edge states as a function of inter-cell coupling (\(C_2\)) at various system sizes (\(M\)). As the system size increases, the value of \(C_2\) at which the admittance gap becomes asymptotically small reduces and approaches the ideal value of \(C_2=C_1=1\) in the infinite-size limit. (b) Spectra of the magnitude of the impedance taken between the two end nodes (i.e., \(a=1\) and \(b=M\)) of the topolectrical chain as a function of \(C_2\) for different system sizes. The boundary impedance peak occurs at smaller (larger) values of \(C_2\) for a larger (shorter) chains. Note that the unit of admittance eigenvalues is in microsiemens (\(\mu\mathrm{S}\)) while the unit of impedance is in mega-ohms (\(\mathrm{M}\Omega\)). Common parameters: \(C_1=1.0 \, \mu\mathrm{F}\), \(C_{\lambda 1}=C_{\lambda 2}=\gamma=0\), \(L_g=10 \, \mu\mathrm{H}\), with resonance frequency \(f_r \approx 167.1 \, \mathrm{kHz}\).}
  \label{fig2}
\end{figure*}

\begin{figure*}[htp!]
  \centering
    \includegraphics[width=0.9 \textwidth]{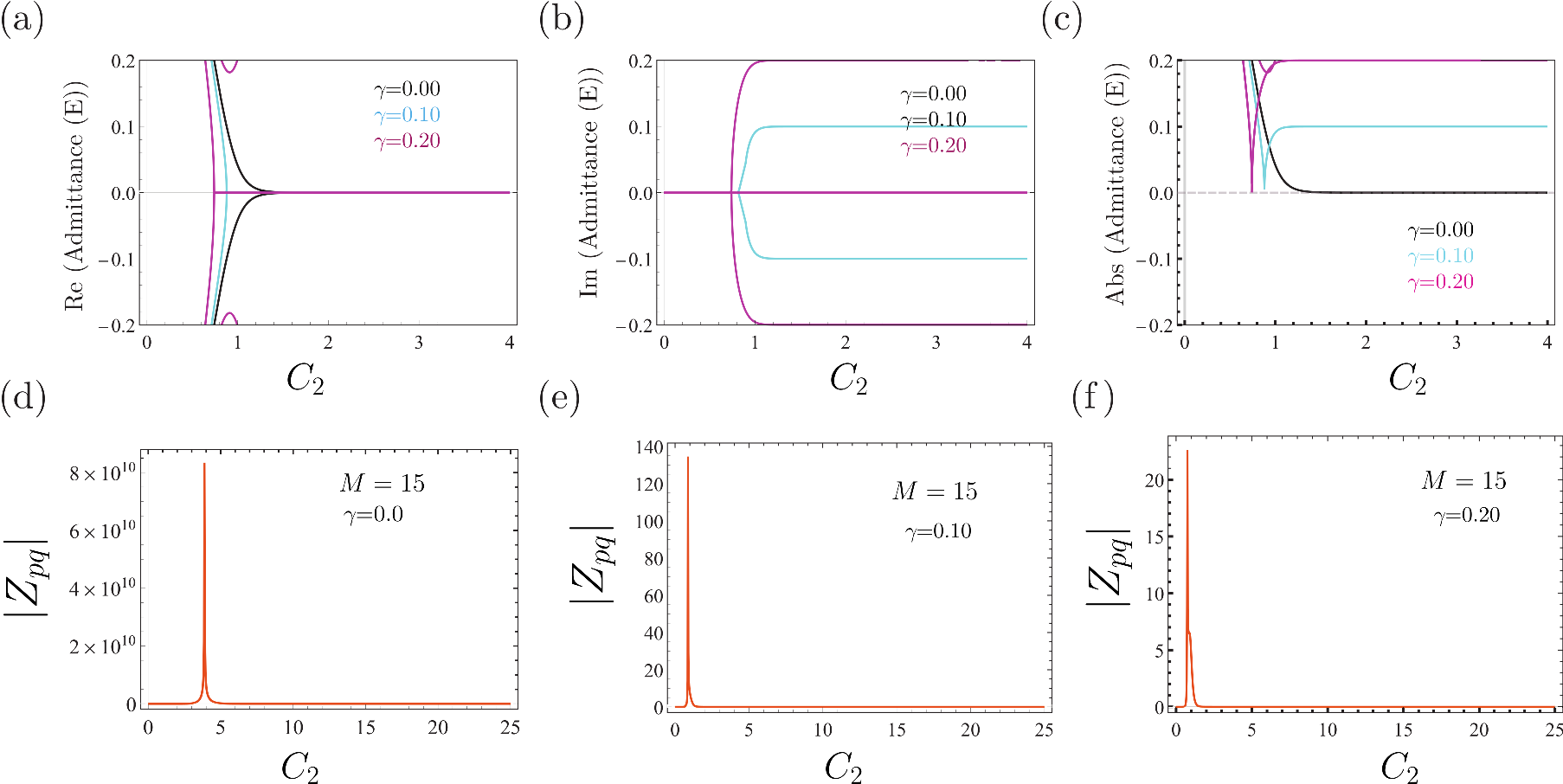}
  \caption{Admittance band structure of the two edge states of the non-Hermitian SSH chain (15 unit cells) at different values of \(\gamma\). The (a) real, (b) imaginary, and (c) absolute values of the admittance of the two edge states as functions of \(C_2\) at various values of \(\gamma\). The TZMs occur at comparatively smaller (larger) values of \(C_2\) at larger (smaller) values of \(\gamma\) and a fixed chain length. The TZMs shift to the ideal value of \(C_2^{\mathrm{ideal}} = \sqrt{C_1^2-C_{\lambda1}^2 +C_{\lambda2}^2}\) when \(\gamma\) is set to a critical value at a given system size. (d-f) The impedance magnitude \(|Z_{pq}|\) between two terminal nodes i.e., \(p=1\) and \(q=30\) at (d) \(\gamma =0.0\), (e) \(\gamma =0.10\), and (f) \(\gamma =0.20\) as functions of \(C_2\). The impedance peaks indicate the location of TZMs, and correspondingly the impedance peak shifts to smaller values of \(C_2\) with the increase of the gain/loss term \(\gamma\). The TZM is located at the \(C_2^{\mathrm{ideal}}\) once \(\gamma\) reaches its critical value. Note that the unit of admittance eigenvalues is in microsiemens (\(\mu\mathrm{S}\)) while the unit of impedance is in mega-ohms (\(\mathrm{M}\Omega\)). Common parameters: \(C_1 = 0.9 \, \mu\mathrm{F}\), \(C_{\lambda 1}=0.1 \, \mu\mathrm{F}\), \(C_{\lambda 2}=0.0 \, \mu\mathrm{F}\), \(L_g=10 \, \mu\mathrm{H}\), \(R_g = 5.0–20.0 \, \mathrm{k}\Omega\) (corresponding to \(\gamma = 0.0–0.2\)), with resonance frequency \(f_r \approx 149.2 \, \mathrm{kHz}\).}
  \label{fig3}
\end{figure*}

This \( f_r \) is independent of \( C_2 \), enabling variation of \( C_2 \) (e.g., in Figs. \ref{fig2} and \ref{fig3}) at a fixed resonance frequency to tune edge states without altering the driving frequency. For notational simplicity, we denote the normalized Laplacian \( (-i \omega)^{-1} \mathcal{J}(\omega) \) at resonant frequency \( f_r \) as \( H(k_x) \), referring to it as the "normalized Laplacian," which is analogous to the lattice Hamiltonian. For a periodic and infinitely large circuit, Bloch's theorem allows us to express the normalized Laplacian matrix, or the equivalent lattice Hamiltonian, in momentum space as
\begin{equation}
H(k_x)= \begin{pmatrix}
i \gamma & (C_1-C_{\lambda1})+(C_2+C_{\lambda2}) e^{-i k_x} \\
(C_1+C_{\lambda1})+(C_2-C_{\lambda2}) e^{i k_x} & -i \gamma
\end{pmatrix}
\label{kHam}
\end{equation}
where \(\gamma=1/(\omega R_g)\) and \(\omega\) is the angular frequency of the driving alternating current. In the absence of staggered gain/loss term (\(\gamma=0\)), the preservation of the chiral symmetry in the TE chain leads to a symmetric admittance spectrum around the zero admittance line (see Supplemental Fig. 1). Under open boundary conditions (OBC), the presence or absence of the topological edge states in the thermodynamic limit signifies which of the two topologically distinct phases the system is in. These phases are associated with different bulk winding numbers of \(|W|=1\) and \(0\), corresponding to the non-trivial and trivial phases when \(|C_2| > |\sqrt{C_1^2-C_{\lambda 1}^2+C_{\lambda2}^2}|\) and \(|C_2| < |\sqrt{C_1^2-C_{\lambda 1}^2+C_{\lambda2}^2}|\), respectively.

However, in finite systems, the edge gap does not close at the exact value of \(C_2 = \sqrt{C_1^2-C_{\lambda 1}^2+C_{\lambda2}^2}\), which occurs for an infinite-sized system. Fig. \ref{fig2} shows the zoomed admittance dispersion (refer to the Appendix and Fig. \ref{figs1} therein for the full admittance dispersions at different system sizes) and impedance spectra of a Hermitian system in which \(C_{\lambda 1}=C_{\lambda 2} = \gamma = 0\). As seen from Fig. \ref{fig2}a, for small system sizes, the edge modes deviate significantly from the zero-admittance line at the ideal infinite-size system value \(C_2=C_1=1\). This phenomenon can be explained by the near-field interaction between edge states \cite{gross2018near}. In contrast, the edge states are located nearer the ideal value of \(C_2=1\) in the longer chain (purple color in Fig. \ref{fig2}a) as the interaction between edge modes becomes weaker. In the Hermitian limit (i.e., \(C_{\lambda1}=C_{\lambda2}=\gamma=0\)), the edge state admittance is given by
\begin{equation}
E_{\mathrm{edge}}=\pm \sqrt{C_1^2+C_2^2-2 C_1 C_2 \cosh \phi}
\label{req2}
\end{equation}
where \(\phi\) satisfies the relation \(e^{\phi}- \xi = \frac{\xi -\xi^3}{1+ \xi^{2M+2}}\) with \(\xi =C_2/C_1\) and \(M\) being the number of unit cells in the finite chain (see detailed derivation in the Methods section). Therefore, at finite system sizes, the ideal zero admittance states of the infinite system split into a pair of nearly-zero energy modes, the eigenvalues of which logarithmically approach zero as \(C_2\) increases beyond the infinite-size critical value of \(C_2=C_1\) (Note that in Fig. \ref{fig2}a-b, the values of \(C_2\) at which the eigenenergies become insignificantly small are approximately 5 and 3 for the system sizes of \(M=25\) and 15, respectively).

This size-dependent edge behavior is also reflected in the impedance measurement. To illustrate this, we plot the impedance spectra measured between the left-most and right-most voltage nodes of the system for different system sizes as a function of \(C_2\) in the Hermitian limit in Fig. \ref{fig2}b. The impedance, defined in Eq. \ref{eqimp}, quantifies the circuit’s response at resonance, where large impedance peaks indicate robust TZMs due to near-zero admittance eigenvalues. The very large impedance peaks appear at different \(C_2\) values (i.e., \(C_2=8\), 4, and 3 for the system sizes of \(M=15\), 20, and 30 unit-cell, respectively) while the theoretical position of the peak in the infinite-size limit is at \(C_2=1\). This discrepancy can be explained via the impedance formula, which reads
\begin{equation}
Z_{pq}=\sum_{l=1}^N \frac{{|\psi_{l,p}-\psi_{k,q}|}^2}{E_l}
\label{eqimp}
\end{equation}
where \(\psi_{l,j}\) is wavefunction amplitude of the \(l^{th}\) eigenmode at the \(j^{th}\) lattice point and \(E_l\) is the \(l^{th}\) non-singular eigenenergy of the Laplacian matrix. Eq. \ref{eqimp} represents the impedance between nodes \(p\) and \(q\), with large values signaling near-zero admittance modes (e.g., TZMs), as the denominator \(E_l\) approaches zero, providing a measurable signature of topological states. Equation \eqref{eqimp} implies that the impedance would assume a very large value whenever the magnitude of one of the eigenvalues approaches zero. Since the admittance eigenvalues approach close to zero at different \(C_2\) values for different system sizes \(M\), the impedance peaks correspondingly occur at different values of \(C_2\) for different \(M\) in Fig. \ref{fig2}. Further analytical discussions on this are given in the Methods section below.

We now turn on both the non-reciprocal coupling (i.e., \(C_{\lambda1} \neq 0\)) and the onsite gain/loss term (i.e., \(\gamma \neq 0\)) to study the effect of non-Hermitian onsite gain/loss on the edge state modulation. The critical system size \(M_c\) (Eq. \ref{req6}) is the minimum number of unit cells where TZMs achieve exactly zero admittance, driven by non-Hermiticity and asymmetric coupling. Below \(M_c\), TZMs are gapped; above \(M_c\), they approach zero energy, yielding large impedance peaks. Based on the Laplacian of Eq. \ref{kHam}, we plot the complex admittance spectra as a function of \(C_2\) at different \(\gamma\) values for a fixed system size \(M\) in Fig. \ref{fig3}a-c (refer to the Appendix and Fig. \ref{figs2} therein for the full admittance dispersions at different values of \(\gamma\)). The presence of non-Hermiticity causes the eigenenergy of the TZM to become exactly zero at some critical system size. As the \(\gamma\) value increases, the exactly zero-energy TZMs shift towards the ideal infinite-system size value of \(|C_2| = |\sqrt{C_1^2-C_{\lambda 1}^2+C_{\lambda2}^2} |\) (see Fig. \ref{fig3}c) which is obtained from the basic SSH model with the similarity transformation to account for the non-reciprocity factors and displacement due to the gain/loss factor. For a system of finite size \(M\), the admittance eigenenergy for the topological edge state can be obtained from the generalized non-Hermitian model and is given by (see the Methods Section III.A for the detailed derivation):
\begin{equation}
E_{\mathrm{edge}}= \pm \sqrt{ -2 \sqrt{(C_1^2-C_{\lambda1}^2)(C_2^2-C_{\lambda2}^2)} \cos \varphi + (C_1^2-C_{\lambda1}^2)+(C_2^2-C_{\lambda2}^2) - \gamma^2 }
\label{req4}
\end{equation}
where \(\varphi =\ln \big( \xi + \frac{\xi -\xi^3}{1+ \xi^{2M+2}} \big)\) with \(\xi=\sqrt{\frac{(C_2^2-C_{\lambda2}^2)}{(C_1^2-C_{\lambda1}^2)}}\). Finally, by setting the above to zero, we find that the size-dependent edge gap vanishes, i.e., the edge states occur at exactly zero admittance when \(C_2\) satisfies
\begin{equation}
\frac{(C_1^2-C_{\lambda1}^2)+(C_2^2-C_{\lambda2}^2) - \gamma^2}{2\sqrt{(C_1^2-C_{\lambda1}^2)(C_2^2-C_{\lambda2}^2)} } = \cos(\varphi)
\label{req5}
\end{equation}
We plot the impedance between the terminal nodes of a finite non-Hermitian SSH chain (15 unit cells long) as a function of \(C_2\) in Fig. \ref{fig3}d--f. The impedance peak exhibits a significant shift as \(\gamma\) is varied, reflecting the dependence of the exceptional points on \(\gamma\), as stated in Eq. \ref{req5}.

\begin{figure*}[htp!]
  \centering
    \includegraphics[width=0.9 \textwidth]{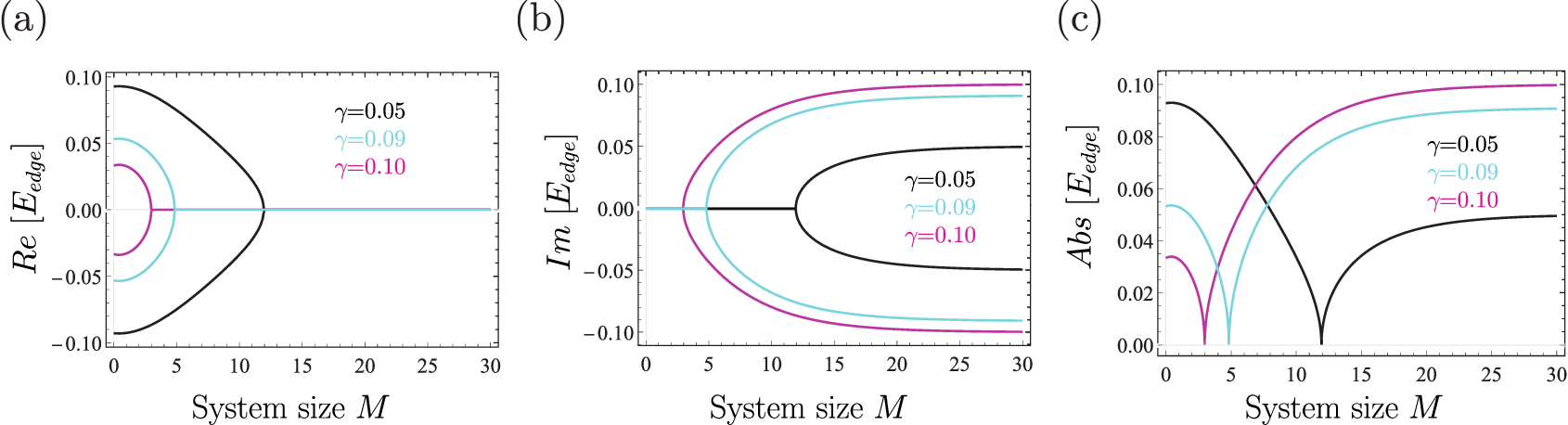}
  \caption{Edge state admittance responses of the non-Hermitian SSH model shown in Fig. \ref{fig1} and their dependence on the size (\(M\)) of the finite lattice. (a) The real, (b) imaginary, and (c) absolute values of two edge modes variation for different \(\gamma\). Increasing \(\gamma\) shifts the position of the exactly zero-energy TZM to a shorter chain. The edge band gap vanishes only at a critical system size (\(M_{\mathrm{c}}\)). Note that the unit of admittance eigenvalues is in microsiemens (\(\mu\mathrm{S}\)). Common parameters: \(C_{1}=0.9 \, \mu\mathrm{F}\), \(C_{\lambda 1}=0.1 \, \mu\mathrm{F}\), \(C_{2}=1.0 \, \mu\mathrm{F}\), \(C_{\lambda 2}=0 \, \mu\mathrm{F}\), \(L_g=10 \, \mu\mathrm{H}\), \(R_g = 5.0–20.0 \, \mathrm{k}\Omega\) (corresponding to \(\gamma = 0.0–0.2\)), with resonance frequency \(f_r \approx 149.2 \, \mathrm{kHz}\).}
  \label{fig4}
\end{figure*}
To clearly illustrate the size dependency of the edge states, we plot the complex admittance spectra as a function of the system size (\(M\)) at three different values of \(\gamma\) in Fig. \ref{fig4}. For a given set of parameters, the critical system size \(M_{\mathrm{c}}\) at which the TZMs occur at exactly zero energy is given by (see the Methods Section III.B for the detailed derivation):
\begin{equation}
M_{\mathrm{c}} = \frac{1}{2} \frac{\ln \big(\frac{2 \xi-\xi^3- e^{\cosh^{-1} \sigma }}{e^{\cosh^{-1} \sigma} -\xi} \big)}{\ln \xi}-1.
\label{req6}
\end{equation}
with \(\sigma = \frac{(C_1^2-C_{\lambda1}^2)+(C_2^2-C_{\lambda2}^2) - \gamma^2}{\sqrt{(C_1^2-C_{\lambda1}^2)(C_2^2-C_{\lambda2}^2)}}\). This is the system size (\(M_{\mathrm{c}}\)) at which the two edge states become fully degenerate with exactly zero admittance value. Moreover, the edge bandgap vanishes completely only when system matches with \(M_{c}\). Note that from Eq. \ref{req6}, the value of \(M_c\), i.e., the system size leading to the existence of TZMs, moves towards a shorter chain with the increase of the gain/loss strength, a trend corroborated in Figs. \ref{fig4}a to \ref{fig4}c.

\begin{figure}[htp!]
  \centering
    \includegraphics[width=0.48 \textwidth]{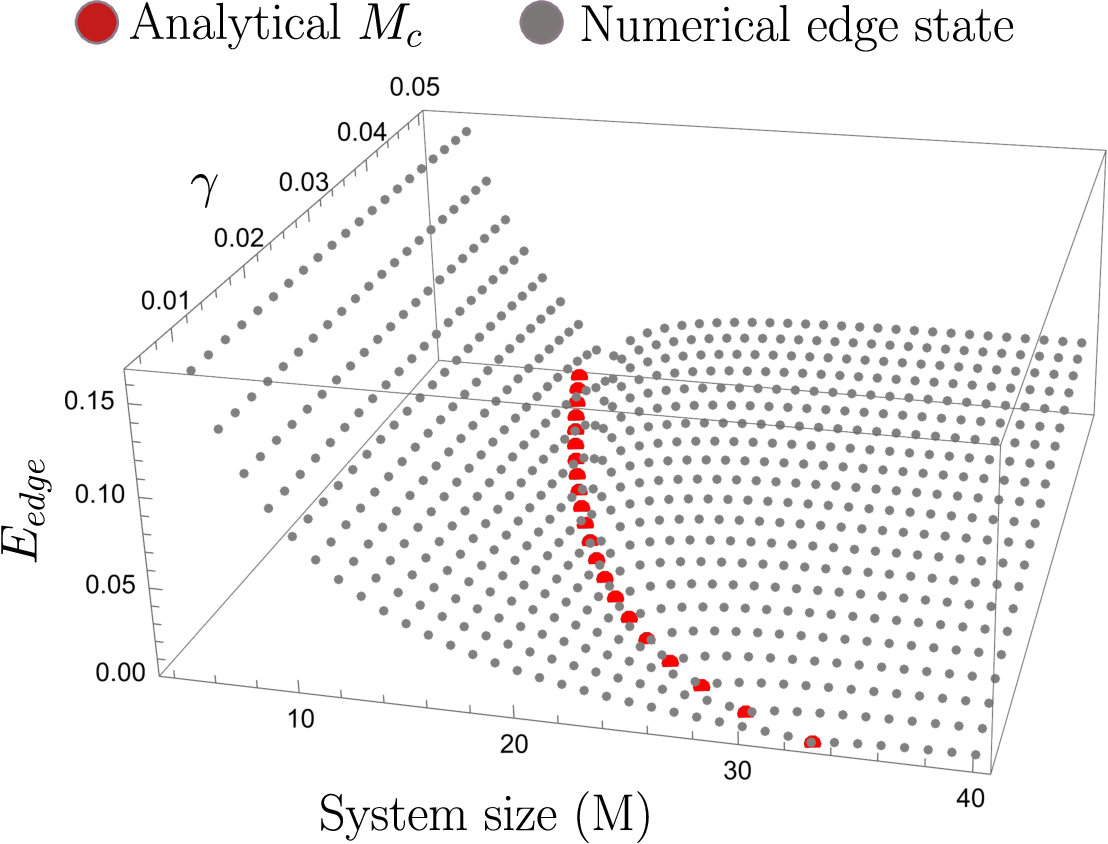}
  \caption{Plot of the theoretical critical system size \(M_c\) (red dots) versus the numerical values of \(M_c\) for different gain and loss parameters \(\gamma\). The red dots show the analytically predicted critical size as given in Eq. \ref{req6}, which decreases with increasing \(\gamma\). The gray dots represent the numerical edge-state energies as functions of system size \(M\) and \(\gamma\). The numerical trend indicates that the edge state energies approach zero energy at smaller \(M\) as \(\gamma\) increases. Common parameters: \(C_{1}=0.9 \, \mu\mathrm{F}\), \(C_{\lambda 1}=0.1 \, \mu\mathrm{F}\), \(C_{2}=1.0 \, \mu\mathrm{F}\), \(C_{\lambda 2}=0 \, \mu\mathrm{F}\), \(L_g=10 \, \mu\mathrm{H}\), \(R_g = 5.0–20.0 \, \mathrm{k}\Omega\) (corresponding to \(\gamma = 0.0–0.2\)), with resonance frequency \(f_r \approx 149.2 \, \mathrm{kHz}\).}
  \label{figMcvsNu}
\end{figure}
To validate the critical system size \( M_c \) derived in Eq.~\(\ref{req6}\), we compare it with the numerically obtained system size at which the edge states acquire exact zero energy. Fig.~\(\ref{figMcvsNu}\) presents a plot of the theoretical predictions against the numerical values of \( M_c \) for different gain and loss parameters \( \gamma \). The red dots represent the analytically predicted critical system size \( M_c \), showing a significant decrease as \( \gamma \) increases. The gray dots correspond to the numerically computed absolute values of the edge-state energies as functions of both \( M \) and \( \gamma \). These results indicate that as the system size increases, the edge states progressively converge to zero energy, and as \( \gamma \) increases, zero-energy edge states emerge at smaller system sizes. The close agreement between the analytical predictions and numerical results confirms the validity of our theoretical model across different gain and loss parameters.

\section{Methods}
\subsection{Derivation of size-dependent edge states}
In this section, we derive the size dependence of the topological edge states. We consider a simple TE circuit for a 1D non-Hermitian SSH chain comprising \(A\) and \(B\) sublattice sites with asymmetric intra and intercell couplings and staggered balanced gain and loss.

The eigenspectrum of the finite-sized chain is identical to that of a tight-binding model with the Hamiltonian
\begin{equation}
\begin{aligned}
H  =&  \sum_{j=1}^{M} \big( (C_1-C_{\lambda1}) a_j^\dag b_j + (C_1+C_{\lambda1}) b_j^\dag a_j + i \gamma (a_j^\dag b_j - b_j^\dag b_j)   \big) \\
& + \sum_{j=1}^{M-1} \big((C_2-C_{\lambda2}) b_j^\dag a_{j+1} + (C_2+C_{\lambda2}) a_j^\dag b_{j+1} \big),
\end{aligned}
\label{Ham1}
\end{equation}
where \(a_j^\dag \) and \(b_j^\dag \) are the creation operator at the \(j\)th unit-cell of sublattice sites \(A\) and \(B\), respectively. Here, \(C_{\lambda \mu}\) (\(\mu=1,2\)) is the distortion factor in the coupling between neighboring nodes that gives rise to asymmetric coupling and \(\gamma\) is half the staggered imaginary potential difference between the A and B nodes. Without loss of generality, we set all couplings parameters to be real and positive.

By definition, an eigenstate \(|\psi\rangle\) of Eq. \eqref{Ham1} with the eigenvalue \(E_{\mathrm{OBC}}\) satisfies \(H |\psi \rangle = | \psi \rangle E_{\mathrm{OBC}}\). This implies that at the A and B nodes of the \(j\)th unit cell,
\begin{equation}
-E_{\mathrm{OBC}} \psi_{(j)\mathrm{A}}+ i \gamma \psi_{(j)\mathrm{A}}+ (C_2+C_{\lambda2}) \psi_{(j-1)\mathrm{B}}+(C_1-C_{\lambda1})\psi_{(j)\mathrm{B}}=0,
\label{eqm2}
\end{equation}
and
\begin{equation}
-E_{\mathrm{OBC}} \psi_{(j)\mathrm{B}}- i \gamma \psi_{(j)\mathrm{B}}+ (C_2-C_{\lambda2}) \psi_{(j+1)\mathrm{A}}+(C_1+C_{\lambda1})\psi_{(j)\mathrm{A}}=0.
\label{eqm3}
\end{equation}
At the two terminal nodes of the finite chain, i.e., the A node at \(j=1\) and the B node at \(j=M\), the KCL equations imply that
\begin{align}
-E_{\mathrm{OBC}} \psi_{(1)\mathrm{A}}+ i \gamma \psi_{(1) \mathrm{A}}+ (C_1-C_{\lambda1})\psi_{(1) \mathrm{B}}&=0  \label{eqm4} \\
-E_{\mathrm{OBC}} \psi_{(M)\mathrm{A}}- i \gamma \psi_{(M) \mathrm{A}}+ (C_1+C_{\lambda1})\psi_{(M) \mathrm{B}}&=0 \label{eqm5}.
\end{align}
Taking the differences between Eq. \eqref{eqm2} at \(j=1\) and Eq. \eqref{eqm4} and between Eq. \eqref{eqm3} at \(j=M\) and \eqref{eqm5} leads to the boundary conditions
\begin{align}
&(C_2+C_{\lambda2})\psi_{(0) \mathrm{B}}=0 \\
&(C_2-C_{\lambda2})\psi_{(M+1)\mathrm{A}}=0
\label{eqm6}.
\end{align}
An eigenstate of Eq. \eqref{Ham1} can generically be written in the form of
\begin{equation}
\begin{pmatrix}\psi_{(j)\mathrm{A}} \\ \psi_{(j)\mathrm{B}} \end{pmatrix} = \sum_{l=1}^2 s_l \beta_l^j \begin{pmatrix} \chi_{\mathrm{A}}^{(l)} \\ \chi_{\mathrm{B}}^{(l)} \end{pmatrix} \label{eqm12}
\end{equation}
where the \(s_l\)s are unknown coefficients to be solved for to satisfy the boundary conditions, and the \(\beta_l\)s and \(\chi_{\mathrm{A/B}}^{(l)}\)s satisfy
\begin{equation}
\begin{pmatrix}
i \gamma & (C_1-C_{\lambda1})+(C_2+C_{\lambda2})\beta_l^{-1} \\
(C_1+C_{\lambda1})+(C_2 - C_{\lambda2})\beta_l  & -i \gamma
\end{pmatrix}\begin{pmatrix}
\chi_{\mathrm{A}}^{(l)} \\ \chi_{\mathrm{B}}^{(l)} \end{pmatrix}= E_{\mathrm{OBC}}\begin{pmatrix}
\chi_{\mathrm{A}}^{(l)} \\ \chi_{\mathrm{B}}^{(l)} \end{pmatrix}.
\label{eqm8}
\end{equation}

Eq. \eqref{eqm8} implies that the \(\beta_l\)s are related to \(E_{\mathrm{OBC}}\) by
\begin{equation}
E_{\mathrm{OBC}}= \pm \sqrt{(C_1+C_{\lambda1})(C_2+C_{\lambda2}) \beta^{-1} + (C_1-C_{\lambda1})(C_2-C_{\lambda2}) \beta + (C_1^2-C_{\lambda1}^2)+(C_2^2-C_{\lambda2}^2)- \gamma^2 }
\label{eqm9}
\end{equation}
where \(\beta\in (\beta_1,\beta_2)\). Solving for \(\beta\) in Eq. \eqref{eqm9} for a given \(E_{\mathrm{OBC}}\), we obtain
\begin{equation}
\beta_1 \beta_2 = \frac{(C_1+C_{\lambda1})(C_2+C_{\lambda2})}{(C_1-C_{\lambda1})(C_2-C_{\lambda2})}.
\label{eqm11}
\end{equation}

Substituting Eq. \eqref{eqm12} into Eq. \eqref{eqm6} gives a system of two linear equations in \(s_1\) and \(s_2\). Because Eq. \ref{eqm8} implies that \(\chi_{\mathrm{A}}^{(l)}\) and \(\chi_{\mathrm{B}}^{(l)}\) are related by
\begin{equation}
\frac{\chi_{\mathrm{A}}^{(l)}}{\chi_{\mathrm{B}}^{(l)}}= \frac{\beta_l (C_1 - C_{\lambda 1}) + C_2 + C_{\lambda 2}}{\beta_l(E_{\mathrm{OBC}}-i\gamma)},
\label{eqm10}
\end{equation}
this system of linear equations can be written in matrix form as
\begin{equation}
\begin{pmatrix}
(C_2+C_{\lambda2})\chi_{\mathrm{B}}^{(1)}& (C_2+C_{\lambda2})\chi_{\mathrm{B}}^{(2)}  \\ (C_2-C_{\lambda2}) \beta_1^{M} (C_2+C_{\lambda2}+(C_1-C_{\lambda1}) \beta_1) \chi_{\mathrm{B}}^{(1)} & (C_2-C_{\lambda2}) \beta_2^{M} (C_2+C_{\lambda2}+(C_1-C_{\lambda1}) \beta_2) \chi_{\mathrm{B}}^{(2)}
\end{pmatrix} \begin{pmatrix}
s_1 \\ s_2 \end{pmatrix}=0.
\label{eqm13}
\end{equation}

A non-trivial equation for \((s_1, s_2)\) exists only when the determinant of the matrix on the left of Eq. \eqref{eqm13} is zero. To simplify the expression for the determinant, we write \(\beta_1= r e^{i \theta}\) and \(\beta_2= r e^{-i \theta}\) where \(\theta\) is, in general, complex except on the complex energy plane GBZ, and \(r\) is the geometric mean of \(\beta_1\) and \(\beta_2\), i.e., the square root of \(\beta_1\beta_2\) given in Eq. \eqref{eqm11}. The condition that the determinant is zero then becomes
\begin{equation}
\sin (M \theta) \big( \xi + \cos \theta \big)+ \cos (M \theta) \sin \theta =0
\label{eqm14}
\end{equation}
where
\begin{equation}
\xi=\sqrt{\frac{C_2^2-C_{\lambda2}^2}{C_1^2-C_{\lambda1}^2}}.
\end{equation}

Rewriting Eq. \eqref{eqm9} in terms of \(\theta\), the bulk admittance eigenvalues for the finite system can be expressed as
\begin{equation}
E^\pm_{\mathrm{OBC}}= \pm \sqrt{ 2 \sqrt{(C_1^2-C_{\lambda1}^2)(C_2^2-C_{\lambda2}^2)} \cos \theta + (C_1^2-C_{\lambda1}^2)+(C_2^2-C_{\lambda2}^2) - \gamma^2 }
\label{eqm15}
\end{equation}

When topological edge states are present, there exists at least one complex solution of \(\theta\) in which \(\theta = \pi+ i \varphi\). In this case, Eq. \eqref{eqm14} becomes
\begin{align}
\frac{\sinh (M+1) \varphi}{\sinh M \varphi} = \xi,
\label{eqm16}
\end{align}
which implies that
\begin{equation}
2 M \varphi = \ln \big( \frac{e^{-\varphi}-\xi}{e^{\varphi}-\xi} \big).
\label{eqm17}
\end{equation}

A solution to Eq. \eqref{eqm17} exists only when the denominator of the fraction on the right-hand side of Eq. \ref{eqm17} is close to 0. We express this by writing \(e^{\varphi} = \xi + \Delta_\varphi\) where \(\Delta_\varphi\) is a small quantity. Eq. \eqref{eqm17} is then transformed into
\begin{equation}
\Delta_\varphi (\xi + \Delta_\varphi)^{2 M}= (\xi + \Delta_\varphi)^{-1} -\xi.
\label{eqm18}
\end{equation}

To estimate \(\Delta_\varphi\), we apply the approximations \((\xi + \Delta_\varphi)^{-1}\approx \frac{\xi-\Delta_\varphi}{\xi^2}\) and \((\xi + \Delta_\varphi)^{2 M} \approx ( \Delta_\varphi)^{2 M}\), which are valid because \(\xi \) is much larger than \(\Delta_\varphi\). \(\Delta_\varphi\) is then approximately given by
\begin{equation}
\Delta_\varphi \approx \frac{\xi (1-\xi^2)}{1+ \xi^{2M+2}}.
\label{eqm19}
\end{equation}
An approximate expression for \(\varphi\) can finally be obtained by using the relation \(e^{\varphi} = \xi+ \Delta_\varphi\):
\begin{equation}
\varphi \approx \ln \left( \xi + \frac{\xi -\xi^3}{1+ \xi^{2M+2}} \right)
\label{eqm20}
\end{equation}
It is clear from Eq. \eqref{eqm20} that the length of the TE circuit affects \(\varphi\), and therefore the eigenvalues of the edge states in Eq. \eqref{eqm15}. The size-dependent edge state eigenvalues can then be expressed in terms of \(\varphi\) as
\begin{equation}
E^\pm_{\mathrm{edge}}= \pm \sqrt{ -2 \sqrt{(C_1^2-C_{\lambda1}^2)(C_2^2-C_{\lambda2}^2)} \cosh \varphi + (C_1^2-C_{\lambda1}^2)+(C_2^2-C_{\lambda2}^2) - \gamma^2 }
\label{eqm21}
\end{equation}
where \(\varphi\) is given by Eq. \eqref{eqm20}.

\subsection{Critical system size for exactly-zero admittance edge modes}
In this sub-section, we will estimate the system size at which the topological edge states are located at exactly zero admittance. To do this, we first calculate the bandgap between two edge states for a finite chain using Eq. \ref{eqm21} as
\begin{equation}
\begin{aligned}
\Delta {E_{\mathrm{edge}}} = & E_{\mathrm{edge}}^{+}-E_{\mathrm{edge}}^{-} \\
&=2 \sqrt{ -2 \sqrt{(C_1^2-C_{\lambda1}^2)(C_2^2-C_{\lambda2}^2)} \cosh \varphi + (C_1^2-C_{\lambda1}^2)+(C_2^2-C_{\lambda2}^2) - \gamma^2 }
\end{aligned}
\label{eqm22}
\end{equation}
Thus, the bandgap between edge states varies significantly with \(\varphi\), which is in turn a function of \(M\) via Eq. \eqref{eqm20}. To find the exact system size \(M=M_{\mathrm{c}}\) at which the edge bandgap vanishes and the edge states become degenerate and pinned at zero admittance energy, we set \(\Delta_{E_{\mathrm{edge}}} = 0\) and solve for the value of \(\varphi\). \(\varphi\) can be expressed as
\begin{equation}
\varphi =\cosh^{-1} \sigma
\label{eqm23}
\end{equation}
where
\begin{equation}
\sigma = \frac{(C_1^2-C_{\lambda1}^2)+(C_2^2-C_{\lambda2}^2) - \gamma^2}{\sqrt{(C_1^2-C_{\lambda1}^2)(C_2^2-C_{\lambda2}^2)}}.
\end{equation}

The critical system size \(M_{\mathrm{c}}\) can then be obtained by equating Eq. \eqref{eqm20} to \eqref{eqm23}, which yields
\begin{equation}
M_{\mathrm{c}} = \frac{1}{2} \frac{\ln \left(\frac{2 \xi-\xi^3- \epsilon}{\epsilon-\xi}\right)}{\ln \xi}-1.
\label{eqm24}
\end{equation}
where \(\epsilon \equiv \exp(\cosh^{-1} \sigma)\).

\subsection{Conclusion}
In this work, we have analyzed a finite non-Hermitian SSH chain model. It is well-known that the classic Hermitian SSH chain of infinite length exhibits topological zero-energy edge modes, the presence or absence of which depends on the ratio of the coupling parameter values. If the Hermitian SSH chain has a finite length, then the degeneracy of the edge modes is lifted resulting in a finite energy gap. Surprisingly, in a finite non-Hermitian SSH model, the topological zero-energy edge modes can be recovered at some critical system size. We provide the exact analytical solution for the energy gap in the finite non-Hermitian SSH chain and the critical size and coupling parameter values at which the topological zero modes emerge. The results hold practical significance due to finite chain lengths and the inevitable imperfections in real-world implementations that lead to non-Hermiticity. We propose a TE implementation of the finite non-Hermitian SSH chain in view of its accessibility. Furthermore, the signature for the presence of the topological zero modes can be easily discerned in a TE platform based on the very large impedance readout between the terminal nodes arising from the vanishing band gap. Additionally, coupling multiple non-Hermitian SSH chains, as explored in prior studies for robust sensor design \cite{Yao2021}, could enhance topological phenomena, such as the formation of bound states, potentially amplifying impedance peaks or stabilizing TZMs for applications in tunable sensors or energy transfer devices. The robustness of TZMs against component tolerances ($\pm 1–5\%$) is analyzed in Appendix (Fig. \ref{fig8}), confirming their practical feasibility with high-precision components. Our results for the size-dependent topological edge states not only provide further insights into the behavior of topological edge states in finite systems but also provide an avenue for designing practical topological systems with tunable boundary modes.

\section{Appendix}

\subsection{Admittance Spectra for Finite Systems}
\begin{figure*}[htp!]
  \centering
    \includegraphics[width=0.9 \textwidth]{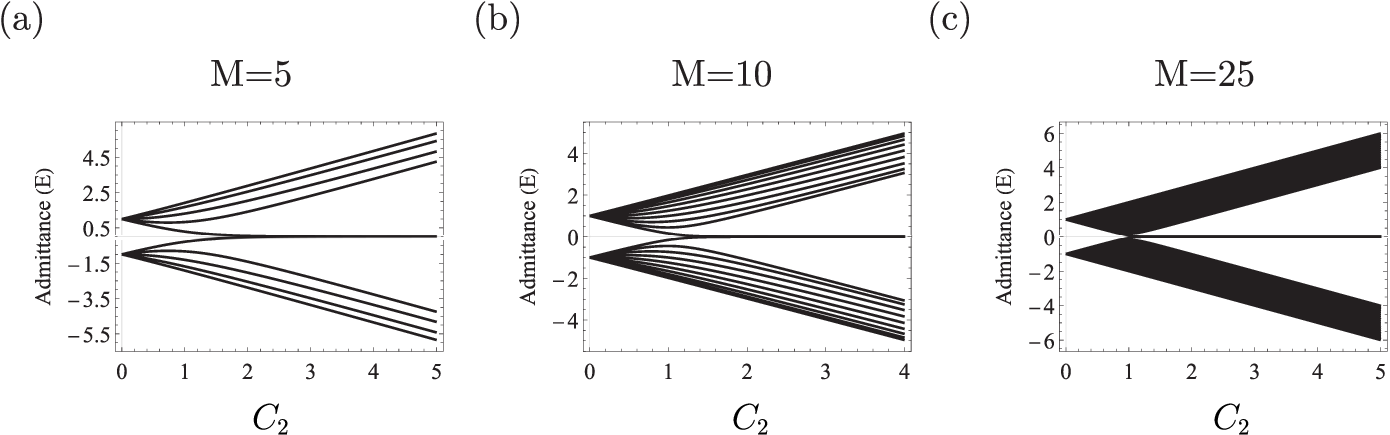}
  \caption{(a-c) Full Admittance spectra (of Fig. \ref{fig2}) as a function of inter-cell coupling (\(C_2\)) for various system size (\(M\)), (a) \(M=5\), (b) \(M=10\), and (c) \(M=25\) unit cells. The bulk modes become denser and the two edge modes become pinned at nearly zero admittance for larger system size. \(C_2=C_1\) is the ideal infinite-size critical value that separates the two topologically distinct regimes. However, in the finite system, the bandgap becomes insignificantly small at different \(C_2\) values that depend on the system size (\(M\)) that decrease and approach \(C_1\) as the system size increases. Note that the unit of admittance eigenvalues is in microsiemens (\(\mu\mathrm{S}\)). Common parameters: \(C_{1}=1.0 \, \mu\mathrm{F}\), \(C_{\lambda 1}=C_{\lambda 2}=0 \, \mu\mathrm{F}\), \(\gamma=0\), \(L_g=10 \, \mu\mathrm{H}\), with resonance frequency \(f_r \approx 167.1 \, \mathrm{kHz}\).}
  \label{figs1}
\end{figure*}

In Fig. \ref{figs1}(a)- (c), we present the full admittance spectra for various system sizes \( M \), depicting the dependence of spectral features on the inter-cell coupling \( C_2 \). Each subfigure corresponds to a different number of unit cells: (a) \( M = 5 \), (b) \( M = 10 \), and (c) \( M = 25 \). As the system size increases, the density of bulk modes also increases, with the two edge modes converging towards zero admittance. The magnified version of the admittance spectra of Fig. \ref{figs1} close to zero admittance value is given in Fig. \ref{fig2} of the main text.

This convergence of edge modes at nearly zero admittance reflects the emergence of topologically protected states. For an ideal infinite system, the critical inter-cell coupling value \( C_2 = C_1 \) marks a topological phase transition, separating distinct topological regimes. However, for finite systems, this transition does not occur precisely at \( C_2 = C_1 \); instead, the bandgap closure point shifts depending on the system size \( M \), approaching \( C_1 \) as \( M \) increases. This size dependence of the bandgap closure highlights finite-size effects, which diminish as the lattice size approaches the infinite limit. For each plot, the parameters \( C_{1} = 1.0 \, \mu\mathrm{F}\), \(C_{\lambda 1}=C_{\lambda 2}=0 \, \mu\mathrm{F}\), \(\gamma = 0\), and \(L_g=10 \, \mu\mathrm{H}\) are held constant, yielding \(f_r \approx 167.1 \, \mathrm{kHz}\), focusing strictly on how \( M \) influences the admittance spectrum and the edge mode behavior.

\begin{figure*}[htp!]
  \centering
    \includegraphics[width=0.9 \textwidth]{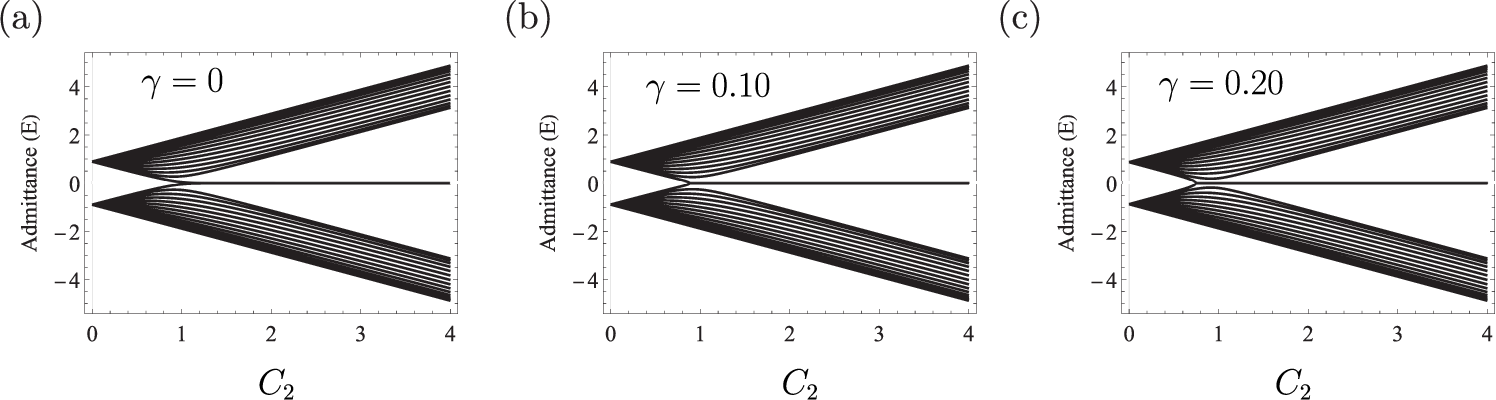}
  \caption{(a-c) Real part of full admittance spectra (of Fig. \ref{fig3}) as function of inter-cell coupling (\(C_2\)) for various \(\gamma\) values at a fixed system size of \(M=15\), (a) \(\gamma=0.0\), (b) \(\gamma=0.10\), and (c) \(\gamma=0.20\). Note that the unit of admittance eigenvalues is in microsiemens (\(\mu\mathrm{S}\)). Common parameters: \(C_1 = 0.9 \, \mu\mathrm{F}\), \(C_{\lambda 1}=0.1 \, \mu\mathrm{F}\), \(C_{\lambda 2}=0.0 \, \mu\mathrm{F}\), \(L_g=10 \, \mu\mathrm{H}\), \(R_g = 5.0–20.0 \, \mathrm{k}\Omega\) (corresponding to \(\gamma = 0.0–0.2\)), with resonance frequency \(f_r \approx 149.2 \, \mathrm{kHz}\).}
  \label{figs2}
\end{figure*}

Fig. \ref{figs2} (a) - (c) illustrate the real part of the full admittance spectra as a function of the inter-cell coupling \( C_2 \) for various values of the gain/loss parameter \( \gamma \) at a fixed system size of \( M = 15 \). Specifically, (a) corresponds to \( \gamma = 0.0 \), (b) to \( \gamma = 0.10 \), and (c) to \( \gamma = 0.20 \). The common parameters for these simulations are \(C_1 = 0.9 \, \mu\mathrm{F}\), \(C_{\lambda 1}=0.1 \, \mu\mathrm{F}\), \(C_{\lambda 2}=0.0 \, \mu\mathrm{F}\), \(L_g=10 \, \mu\mathrm{H}\), \(R_g = 5.0–20.0 \, \mathrm{k}\Omega\) (corresponding to \(\gamma = 0.0–0.2\)), with resonance frequency \(f_r \approx 149.2 \, \mathrm{kHz}\). Here, we introduce both non-reciprocal coupling (i.e., \( C_{\lambda 1}, C_{\lambda 2} \neq 0 \)) and the onsite gain/loss term (i.e., \( \gamma \neq 0 \)). The magnified version of the admittance spectra of Fig. \ref{figs2} close to zero admittance value is given in Fig. \ref{fig3} of the main text. The complex admittance spectra are plotted as a function of \( C_2 \) across the different \( \gamma \) values. Notably, the introduction of non-Hermiticity alters the eigenenergies of the topological zero modes (TZMs), causing them to shift to exactly zero at a specific system size. As the value of \( \gamma \) increases, these zero-energy TZMs move closer to the ideal infinite-system size value given by

\[
|C_2| = |\sqrt{C_1^2 - C_{\lambda 1}^2 + C_{\lambda 2}^2} |,
\]
as depicted in details in Fig. \ref{fig3}(c). This shift underscores the significant influence of gain/loss terms on the behavior of topological modes in finite systems, providing insights into the non-Hermitian effects on the admittance spectra.

\subsection{Experimental Feasibility}
To facilitate experimental verification of the non-Hermitian SSH TE circuit (Fig. \ref{fig1}), we propose realistic parameter values based on prior TE circuit studies \cite{helbig2020generalized,rafi2021topological,zou2021observation}. Capacitances are set as \(C_1 = 0.9 \, \mu\mathrm{F}\), \(C_2 = 0.5–1.5 \, \mu\mathrm{F}\), \(C_{\lambda 1} = 0.1 \, \mu\mathrm{F}\), and \(C_{\lambda 2} = 0.0 \, \mu\mathrm{F}\), using standard ceramic capacitors and negative impedance converters (NICs) to implement \(-C_2\). The grounding inductor is \(L_g = 10 \, \mu\mathrm{H}\), yielding a resonance frequency \(f_r \approx 149.2 \, \mathrm{kHz}\), calculated via \(f_r = \frac{1}{2 \pi \sqrt{L_g (C_1 + C_{\lambda 1} + C_{\lambda 2})}}\). The tunable resistor \(R_g = 5.0–20.0 \, \mathrm{k}\Omega\) achieves \(\gamma = 0.0–0.2\), implementable via variable resistors or NICs with operational amplifiers (e.g., LM358). A circuit with 15–25 unit cells, matching the critical size \(M_c\) (Eq. \ref{req6}), can be built on a printed circuit board and measured with an LCR meter to detect TZM impedance peaks (Fig. \ref{fig3}d–f).

\subsection{Impact of Component Tolerances on Edge States}
\begin{figure*}[htp!]
  \centering
    \includegraphics[width=0.8 \textwidth]{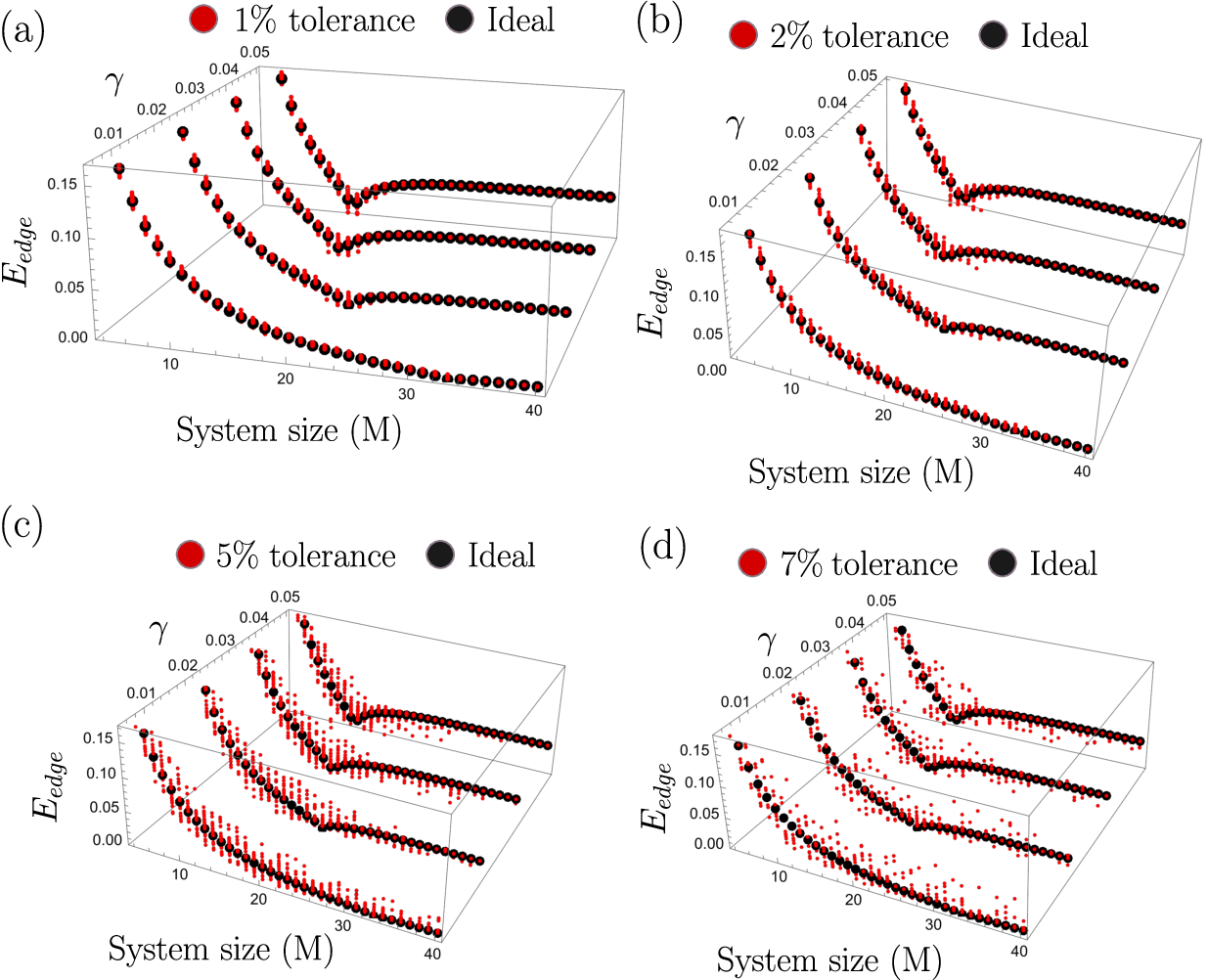}
  \caption{Absolute values of edge state energies as functions of system size \( M \) and gain/loss parameter \( \gamma \), showing the impact of component tolerances. (a) ±1\% tolerance, (b) ±2\% tolerance, (c) ±5\% tolerance, and (d) ±7\% tolerance. Black dots represent the ideal case (no tolerances), while red dots show the effect of tolerances on capacitors (\( C_1 \), \( C_2 \), \( C_{\lambda 1} \), \( C_{\lambda 2} \)) and resistors (\( R_g \)). At low tolerances (a–b), edge states closely match the ideal case; at higher tolerances (c–d), slight shifts occur, affecting the critical size \( M_c \) by ±1–2 unit cells. Common parameters: \( C_1 = 0.9 \, \mu\mathrm{F}\), \( C_{\lambda 1} = 0.1 \, \mu\mathrm{F}\), \( C_2 = 1.0 \, \mu\mathrm{F}\), \( C_{\lambda 2} = 0 \, \mu\mathrm{F}\), \(L_g=10 \, \mu\mathrm{H}\), \(R_g = 5.0–20.0 \, \mathrm{k}\Omega\) (corresponding to \(\gamma = 0.0–0.2\)), with resonance frequency \(f_r \approx 149.2 \, \mathrm{kHz}\).}
  \label{fig8}
\end{figure*}

To assess the robustness of the topological zero modes (TZMs) against manufacturing tolerances in real-world TE circuits, we analyze the effect of component tolerances on the critical system size \( M_c \) (Eq. \ref{eqm24}). Fig. \ref{fig8} illustrates the absolute values of edge state energies as functions of system size \( M \) and gain/loss parameter \( \gamma \), comparing the ideal case (black dots) with cases including component tolerances (red dots) for capacitors (\( C_1 \), \( C_2 \), \( C_{\lambda 1} \), \( C_{\lambda 2} \)) and resistors (\( R_g \)). Subfigures \ref{fig8}(a) and (b) show tolerances of ±1\% and ±2\%, respectively, where edge state energies closely align with the ideal case, indicating minimal impact on \( M_c \). Subfigures Fig. \ref{fig8}(c) and (d) show higher tolerances of ±5\% and ±7\%, respectively, where slight shifts in the edge state profiles occur, altering \( M_c \) by approximately ±1–2 unit cells. These shifts, driven by variations in coupling ratios (\( C_1/C_2 \), \( C_{\lambda 1}/C_{\lambda 2} \)) and \( \gamma \), do not eliminate TZMs but may narrow the zero-mode resonance. High-precision components (±1\%) can mitigate these effects, ensuring robust TZM detection. Common parameters are \( C_1 = 0.9 \, \mu\mathrm{F}\), \( C_{\lambda 1} = 0.1 \, \mu\mathrm{F}\), \( C_2 = 1.0 \, \mu\mathrm{F}\), \( C_{\lambda 2} = 0 \, \mu\mathrm{F}\), \(L_g=10 \, \mu\mathrm{H}\), \(R_g = 5.0–20.0 \, \mathrm{k}\Omega\) (corresponding to \(\gamma = 0.0–0.2\)), with resonance frequency \(f_r \approx 149.2 \, \mathrm{kHz}\).

\subsection{Non-Hermitian Skin Effect and Eigenstate Localization}\label{subsec:nhse}

\begin{figure*}[htp!]
  \centering
    \includegraphics[width=0.6 \textwidth]{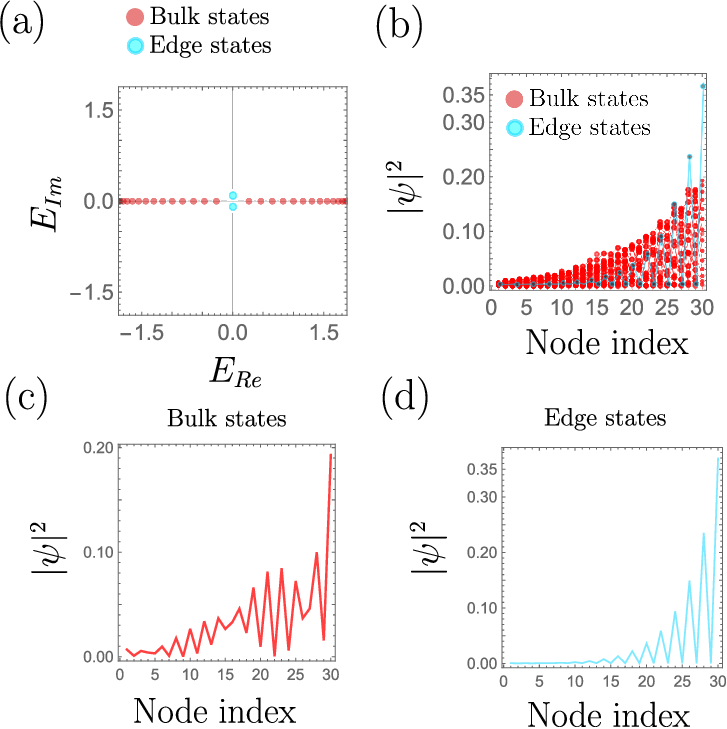}
  \caption{(a) Complex admittance eigenvalue spectrum for a non-Hermitian SSH chain with \(M = 15\) unit cells, at fixed \(C_2 = 1.0 \, \mu\mathrm{F}\) and resonant frequency \(f_r \approx 149.2 \, \mathrm{kHz}\), showing edge modes in the bandgap near zero admittance and dense bulk modes. (b) Eigenstates distribution of the circuit model. Bulk and edge states are denoted by red and cyan colors, respectively. (c) Eigenstate distribution of a representative bulk mode, showing slight asymmetric localization towards one boundary due to a weak non-Hermitian skin effect (NHSE), characterized by the non-Bloch factor \(\beta \approx 1.118\). (d) Eigenstate distribution of a representative edge mode, localized at the boundaries, indicative of topological zero modes (TZMs). Common parameters: \(C_1 = 0.9 \, \mu\mathrm{F}\), \(C_{\lambda 1}=0.1 \, \mu\mathrm{F}\), \(C_{\lambda 2}=0.0 \, \mu\mathrm{F}\), \(C_2 = 1.0 \, \mu\mathrm{F}\), \(L_g = 10 \, \mu\mathrm{H}\), \(\gamma = 0.1\), \(R_g = 10 \, \mathrm{k}\Omega\).}
  \label{figs3}
\end{figure*}

In non-Hermitian systems with asymmetric coupling, such as our topolectrical SSH circuit with non-reciprocal capacitances \(C_{\lambda 1} \neq 0, \), the non-Hermitian skin effect (NHSE) can lead to the localization of bulk eigenstates at one boundary, significantly altering the bulk-boundary correspondence \cite{PhysRevResearch.4.043108}. To investigate the presence of NHSE in our model, we analyze the admittance eigenvalue spectrum and eigenstate distributions under open boundary conditions (OBC) at a fixed inter-cell coupling \(C_2 = 1.0 \, \mu\mathrm{F}\) and resonant frequency \(f_r \approx 149.2 \, \mathrm{kHz}\), determined by \(f_r = \frac{1}{2 \pi \sqrt{L_g (C_1 + C_{\lambda 1} + C_{\lambda 2})}}\) with \(C_1 = 0.9 \, \mu\mathrm{F}\), \(C_{\lambda 1} = 0.1 \, \mu\mathrm{F}\), \(C_{\lambda 2} = 0.0 \, \mu\mathrm{F}\), and \(L_g = 10 \, \mu\mathrm{H}\). Figure \ref{figs3} presents the admittance eigenvalue spectrum and corresponding eigenstate distributions for a system with \(M = 15\) unit cells, highlighting the interplay between topological zero modes (TZMs) and potential NHSE-induced bulk mode localization.

The Hamiltonian under periodic boundary conditions (PBC), given in Eq. \ref{kHam}, is:
\[
H(k_x) = \begin{pmatrix}
i \gamma & (C_1 - C_{\lambda 1}) + (C_2 + C_{\lambda 2}) e^{-i k_x} \\
(C_1 + C_{\lambda 1}) + (C_2 - C_{\lambda 2}) e^{i k_x} & -i \gamma
\end{pmatrix},
\]
where \(\gamma = 1/(\omega R_g)\) represents the gain/loss term, and \(\omega = 2\pi f_r\). The NHSE arises due to asymmetric coupling, characterized by the non-Bloch factor \(\beta\), which replaces \(e^{i k_x}\) in the generalized Brillouin zone (GBZ) for OBC \cite{PhysRevResearch.4.043108}. To derive \(\beta\), we consider the eigenvalue equation for the Hamiltonian under OBC, where the eigenstates take the form \(\psi_j = \beta^j \begin{pmatrix} \chi_A \\ \chi_B \end{pmatrix}\). The characteristic equation for the eigenvalues \(E\) is obtained by setting the determinant of \(H(\beta) - E I = 0\):
\[
E^2 + 2i\gamma E - \left[ (C_1 - C_{\lambda 1})(C_1 + C_{\lambda 1}) + (C_2 + C_{\lambda 2})(C_2 - C_{\lambda 2}) \beta + (C_1 + C_{\lambda 1})(C_2 + C_{\lambda 2}) \beta^{-1} \right] + \gamma^2 = 0.
\]
The non-Bloch factor \(\beta\) is determined by the condition that the bulk eigenstates localize under OBC. For simplicity, consider the case with \(\gamma = 0\) (no gain/loss) and \(C_{\lambda 2} = 0\), as in our parameter regime. The characteristic equation simplifies, and the localization factor \(\beta\) is found by solving the quadratic equation in \(\beta\):
\[
(C_2)(C_1 + C_{\lambda 1}) \beta^{-1} + (C_1 - C_{\lambda 1})(C_2) \beta = -(C_1^2 - C_{\lambda 1}^2).
\]
Multiplying through by \(\beta\), we obtain:
\[
(C_1 + C_{\lambda 1})(C_2) + (C_1 - C_{\lambda 1})(C_2) \beta^2 = -(C_1^2 - C_{\lambda 1}^2) \beta,
\]
yielding the non-Bloch factor:
\[
\beta^2 = \frac{(C_1 + C_{\lambda 1})(C_2)}{(C_1 - C_{\lambda 1})(C_2)} = \frac{C_1 + C_{\lambda 1}}{C_1 - C_{\lambda 1}}.
\]
Thus, \(\beta = \sqrt{\frac{C_1 + C_{\lambda 1}}{C_1 - C_{\lambda 1}}}\). For our parameters (\(C_1 = 0.9 \, \mu\mathrm{F}\), \(C_{\lambda 1} = 0.1 \, \mu\mathrm{F}\)), \(\beta = \sqrt{\frac{0.9 + 0.1}{0.9 - 0.1}} = \sqrt{\frac{1.0}{0.8}} \approx 1.118\). The decay rate of bulk eigenstates under NHSE is given by \(\kappa = |\ln |\beta|| \approx |\ln 1.118| \approx 0.112\), indicating weak localization towards one boundary, as \(|\beta|\) is close to 1.

Figure \ref{figs3}(a) shows the complex admittance eigenvalue spectrum for \(M = 15\), \(\gamma = 0.1\), and \(R_g = 10 \, \mathrm{k}\Omega\), with two edge modes within the bandgap near zero admittance and dense bulk modes. Figure \ref{figs3}(b–d) displays the eigenstate distributions for a representative edge mode and a bulk mode, respectively. The edge mode is strongly localized at the boundaries, consistent with TZMs, while the bulk mode shows a slight asymmetry towards one boundary, indicative of a weak NHSE due to the modest \(\beta \approx 1.118\). This weak NHSE does not dominate the TZM recovery, which is our primary focus, as the circuit design with the \(-C_2\) capacitor prioritizes size-dependent edge mode behavior at a fixed \(f_r\). Common parameters are \(C_1 = 0.9 \, \mu\mathrm{F}\), \(C_{\lambda 1} = 0.1 \, \mu\mathrm{F}\), \(C_{\lambda 2} = 0.0 \, \mu\mathrm{F}\), \(C_2 = 1.0 \, \mu\mathrm{F}\), \(L_g = 10 \, \mu\mathrm{H}\), with \(f_r \approx 149.2 \, \mathrm{kHz}\).

\subsubsection*{Acknowledgements}
This work is supported by the Ministry of Education (MOE) Tier-II Grant MOE-T2EP50121-0014 (NUS Grant No. A-8000086-01-00), and MOE Tier-I FRC Grant (NUS Grant No. A-8000195-01-00).


%

\end{document}